# IDENTITY MANAGEMENT ON BLOCKCHAIN – PRIVACY AND SECURITY ASPECTS

## (EXTENDED VERSION[1])

Andreea-Elena PANAIT[*], Ruxandra F. OLIMID[*,**,***], Alin STEFANESCU[*,**]

[*] Department of Computer Science, University of Bucharest, Romania
[**] The Research Institute of the University of Bucharest (ICUB), Romania
[***] Department of Information Security and Communication Technology,
NTNU - Norwegian University of Science and Technology, Trondheim, Norway
E-mail: `andreea-elena.panait@drd.unibuc.ro`, `{ruxandra.olimid,alin}@fmi.unibuc.ro`

**Abstract.** In the last years, identity management solutions on blockchain were proposed as a possible solution to the digital identity management problem. However, they are still at an early stage and further research needs to be done to conclude whether identity systems could benefit from the use of blockchain or not. Motivated by this, we investigate identity management solutions on blockchain intending to give the reader an overview of the current status and provide a better understanding of the pros and cons of using such solutions. We conduct an analysis on ten of the most known implementations, with a focus on privacy and security aspects. Finally, we identify existing challenges and give new directions for research.

## 1. INTRODUCTION

### 1.1 Motivation and contribution

Identity management became one of the major problems in today's world. Because of the continuous technological development, specifically the development of 5G and Internet of Things (IoT), the number of entities in the digital world increased significantly. In consequence, it is a high demand to digitally identify not only individuals, but also organizations, services, applications, and devices in an efficient and interoperable (or ideally, universal) way. In the view of international privacy rights regulations, such as the General Data Protection Regulation (GDPR) [1], special attention must be given to privacy and security aspects. Data leaks or identity fraud (e.g., stealing of credentials or credit card information) often result in economic loss and decrease of trust in the identity providers. Within this context, new identity management solutions that satisfy the actual needs must be explored. Identity management solutions based on blockchain technology were proposed as a possible solution to solve these needs. In the last two years, studies show that about 50% of blockchain service providers are targeting the use case of digital identity [2, 3]. Lately, big companies such as Facebook, are analyzing the possibility of integrating blockchain technology into the platform's data-sharing systems and login [4]. Moreover, Facebook recently founded a new association,

---

[1] Original paper in Proceedings of the Romanian Academy, Series A, Vol 21, No 1/2020, pp. 45–52:
https://acad.ro/sectii2002/proceedings/doc2020-1/06-Panait.pdf

Libra, through which they plan to bring a new cryptocurrency and blockchain platform to be used by billions of people. One of the prerequisites for that is a new identity management based on blockchain. As they put it in [5], *"An additional goal of the association is to develop and promote an open identity standard. We believe that decentralized and portable digital identity is a prerequisite to financial inclusion and competition."*

Nevertheless, despite the great interest in blockchain in both academia and production, blockchain-based identity management solutions are still at an early stage. While some tend to erroneously consider blockchain a solution to many problems, there have been lately a lot of discussions about whether identity systems could benefit from blockchain or not. Motivated by this, we investigate identity management solutions on blockchain.

The goal of the paper is to provide an overview of the current status in the field and a better understanding of the pros and cons of using such solutions. Furthermore, we conduct an analysis on ten of the most known implementations, with a focus on privacy and security aspects, which is the main contribution of the paper. Although some similar work has been done in the area, existing papers either refer to just a few specific solutions [6, 7] or consider only specific models or use-cases [8, 9, 10]. Another contribution is our discussion on the limitations and weaknesses of identity management on blockchain, identification of current challenges and new directions of research.

The paper is organized as follows. The next subsection introduces the related work. Section 2 gives the necessary background in digital identity and blockchain. Section 3 presents identity management on blockchain, with a focus on particularities, classification and investigates existing implementations. Section 4 discusses the necessities and challenges of blockchain-based identity management with a highlight on open research questions. Finally, we conclude.

**1.2 Related work**

User-centric identity management models have been adopted for years now [11], in solutions such as *OpenID* [12]. Recently, a great amount of work has been done for designing identity management on blockchain and closely related fields, such as public-key infrastructure [13, 14]. Mühle et al. give an overview of self-sovereign identity, with a focus on general architecture, authentication solutions to link a user to his digital identifier, and verifiable claims [9]. Dunphy et al. discuss the role of the distributed ledger technology in the context of digital identity, identify challenges, and propose future research directions [15]. Zhu and Badr analyze digital identity management in the context of IoT, with a focus on self-sovereign solutions [10]. Bokkem et al. [16] explore the self-sovereign solutions available and discuss their implementations with respect to the principles previously described by C. Allen [17]. Dunphy and Petitcolas offer a view on the distributed ledger technology-based identity management and evaluate three system proposals (*uPort*, *ShoCard,* and *Sovrin*) [6]. Soltani et al. present an identity management framework for banking called KYC2 which is based on self-sovereign principles and the Hyperledger Indy (a permissioned public distributed ledger) [18]. Roos explains the design of some of the most promising identity management solutions that use blockchain and evaluates whether the solutions fulfill their goals or not [7]. Many other solutions have been published in the last years. From these, we only mention a few: Hardjono and Pentland described *ChainAnchor* [19], Liu et al. introduced an identity management solution based on Ethereum smart contracts [20], Augot et al. proposed a solution on Bitcoin making use of zero-knowledge proofs (ZKP) [21], Faber et al. propose *BPDIMS*, a human-centric approach [22]. The implemented solutions *Namecoin* [23], *Blockstack* [24], *uPort* [25], *Sovrin* [26], *EverID* [27], *SelfKey* [28], *ShoCard* [29], *Sora* [30], *LifeID* [31], and *IDchainZ* [32] will be discussed in more details in Subsection 3.3.

## 2. BACKGROUND

### 2.1 Digital Identity

A *digital identity* is a representation of an actual entity (e.g., a person, a device, an organization) in the digital world. A digital identity is not exhaustive (i.e., only part of the actual identity is comprised of the attributes bounded to a digital identity), and it can vary depending on the specific domain it is used for. An entity can have several digital identities, used in different contexts for distinct purposes. However, digital identifiers should be unique, in the sense that there must not be two equal identifiers for distinct entities. This would result in the impossibility of identification. The attributes of a digital identity are e.g., physical characteristics of a person, address, e-mail, phone number [33]. Similar to attributes of an actual identity, some of them might change over time. From a technical point of view, credentials (username and password), security tokens, or transaction history can also be used for identification. On the other hand, individuals can also be uniquely identified by single-use identifiers or pseudonymous especially created for different purposes or specific intervals of time [34, 35].

A *digital identity management model* can be traditionally interpreted as a three-party model that consists of an *end-user* (the actual entity that possesses a digital identity and wants to perform an action), *an identity provider* (the entity that enrolls new users, manages digital entities and performs authentication), and *a service provider* (or *relying party*, the entity that provides services to the end-user, and it relies on the identity provider to verify the identity of the end-user) [10, 15, 36].

In the literature, there are four major identity management models: (1) *the isolated identity model*, (2) *the centralized identity model*, (3) *the federated identity model*, and (4) *the user-centric identity model*. In the *isolated identity model*, the service provider is also managing user identity, meaning that the end-user has one digital identity for each service provider [36, 37]. In the *centralized identity model*, the identity provider manages and centralizes digital identities, and the users are authenticated prior to connecting to the service provider; this means that the user can reuse the same credentials for multiple service providers by e.g., *Single Sign-On* (SSO) [36, 37]. In the *federated identity model*, the user can use the same credentials for authenticating and accessing service providers that together with the identity provider form a federation; contrary to the centralized identity model, it makes use of *pseudonyms* (distinct pseudonyms are used for different service providers) [36, 38]. In the *user-centric identity model*, the user has control of his attributes, and he can define his own policies for sharing his identity with the service providers. However, despite the users' capabilities in selecting and change contractual terms, it still relies on identity providers [36, 8]. A more recent model, the *Self-Sovereign Identity (SSI) model* takes a step further from the user-centric model and eliminates the need for an external identity provider. The end-user gains full control of its own identity, being his own identity provider. This eliminates the need for an external authority and, as a consequence, it decreases the danger of identity theft [8, 6]. We will refer more to the SSI model in Subsection 3.2.

Regardless of the model, identity management solutions should satisfy the *laws of identity*, which is an evaluation framework used to identify achievements and faults of digital identity solutions [39]. The laws suggest that identity-related information should be exposed to legitimate parties only (*justifiable parties*), and with the user's approval (*user control and consent*). Moreover, the gathered and stored information should be minimal, in accordance with service needs (*minimal disclosure for a constrained use*). The end-users should understand how to interact with the system and be clearly aware of the implications of their action (*human integration*). They should have the possibility to share identity information in either a private or a public way (*directed identity*). An identity management solution should provide interaction between multiple identity credentials and schemes (*design for pluralism of operators and technology*), and end-users should have a consistent experience regardless of the technology platform and context of security (*consistent experience across contexts*).

Legally, identity management solutions must satisfy privacy and security protections in accordance with the regulations in force. For example, solutions that are implemented in the countries of the European Union must comply with the GDPR, which is a set of policies on data protection that came into force in May 2018 [1, 40]. GDPR states in Article 5 the principles related to the processing of personal data (*lawfulness, fairness and transparency, purpose limitation, data minimization, accuracy, storage limitation, integrity and confidentiality, accountability*). Chapter 3 of GDPR presents the rights of the data subject, and Chapter 4, Section 2 introduces aspects on the security of personal data [1]. Similar principles exist outside Europe as well. For example, the Digital ID & Authentication Council of Canada (DIACC) introduced ten principles that a digital identity ecosystem should follow [41].

## 2.2 Blockchain

A blockchain is a type of decentralized distributed database or a ledger. Sometimes the notions of *Distributed Ledger Technology* (DLT) and blockchain are erroneously used interchangeably. In fact, they are different, a blockchain is a specific implementation of DLT with cryptographic enhancements [6, 18]. A blockchain is *distributed*, in the sense that it is spread between multiple nodes and each node stores a copy of the blockchain. A node does store the latest transactions but does not necessarily maintain the whole history. The nodes that store the entire blockchain are called *full nodes*. A blockchain is also *decentralized* by design in the sense that there is not a single point of decision, but the decision is a result of a *consensus* of the nodes. There are different types of consensus protocols (e.g., Proof of Work (PoW), proof of stake) but their goal is always the same: to decide how the nodes agree on blocks that are validated and added to the blockchain.

Asymmetric cryptography is intensively used in blockchains. Unlike symmetric cryptography where a shared key is used, asymmetric cryptography uses a public-private key pair. Usually, the public key corresponds to the public address of the user in the blockchain. This allows the user to sign a transaction under his private key such that everyone can verify its authenticity in the blockchain (as corresponding to his address). The private key is thus used to authorize actions on the user account, this being a simple method of authentication.

Table 1. Blockchain classification

| | Permissionless | Permissioned |
|---|---|---|
| **Type of nodes** | Public (i.e., open to any node) | Private (i.e., open to nodes that are authorized by the authority that manages the blockchain) |
| **Publish transactions / blocks** | Public | Private |
| **Read transactions / blocks** | Public | Public **or** Private |
| **Data visibility / availability** | Public | Public **or** Private |
| **No. of accepted nodes** | Large and easily scalable | Low and usually not scalable for many nodes |
| **Time to join the network** | Faster to join the network (no authorization / registration) | Slower to join the network (authorization / registration required) |
| **Governance** | Publishing nodes, software developers | The owner / consortium of the blockchain |
| **Consensus protocol** | Usually slower and more expensive (in computational power and resources in general) | Usually faster and less expensive (in computational power and resources in general) |
| **Software** | Open-source, freely available for download | Open-source **or** Closed-source |
| **Malicious nodes and majority domination** | More predisposed to malicious nodes and 51% attack (e.g. Sybil attack) | Less predisposed to malicious nodes and 51% attack (legal measures can be taken against malicious nodes) but directly vulnerable to the owner of the blockchain, which has power to replace / change the blockchain blocks |
| **Conflicts / Forks (resolved by consensus protocols)** | More predisposed to conflicts / forks | Less predisposed to conflicts / forks |
| **End of life** | Difficult (nodes might continue to run) | Easy (legal measures might be taken against nodes still running) |

Blockchains can be classified into *permissionless* and *permissioned.* In a permissionless blockchain, any entity can become a node and participate in the blockchain consensus (e.g., add transactions). On the contrary, permissioned blockchain increase control by limiting participation to specific accepted entities only [42]. Usually, a permissioned blockchain restricts the nodes that participate in the consensus protocol to special sets of nodes. Depending on the application, one type or the other can be used. To exemplify, cryptocurrencies such as *Bitcoin* are permissionless, while enterprise solutions used inside organizations are usually permissioned. Table 1 summarizes some of the properties of permissioned and permissionless blockchains. By *public*, we understand any node that wishes to perform the given action, while by *private* we understand special nodes that are accepted by the owner of the permissioned blockchain to perform that action.

## 3. IDENTITY MANAGEMENT ON BLOCKCHAIN

We now discuss identity management on blockchain, which comes as a candidate solution to the limitations of centralized architectures, in the light of the rapid development of blockchain technologies.

### 3.1 Specific Characteristics of Identity Management on Blockchain

Identity management solutions built on blockchain benefit of the intrinsic advantages of the blockchain technology. They eliminate the need for a central authority to control and manage the system, and they move the responsibility to the user. These advantages are expected to solve (to some extent) some of the problems that appear in centralized systems, such as identity fraud or data leaks [15]. However, this is debatable because identity management solutions either do not achieve full decentralization, or they require a high level of trust [6]. By construction, blockchain brings transparency in data changes, and data history cannot be otherwise manipulated (unless the majority of nodes agree with respect to the change). On the other hand, it introduces challenges in terms of implementation efficiency and even security. We will refer to these in more detail in Section 4.

A blockchain-based identity management solution should allow selective storing of identities in the blockchain. Identities need to be *attested* by authorities or other entities in the blockchain. This usually works as follows. An entity claims an identity by a *verifiable claim*, that is attested after the verification of some *attributes* that differentiate the user (e.g., phone number, e-mail, governmental identity documents (IDs), biometrics). Within identity management on blockchain it is a clear difference between the *digital identifier* (a value that uniquely identifies the entity) and the *attributes* that are associated with it [6]. As a non-authorized or uncontrolled disclosure of attributes results in security and privacy leaks, so attributes storage (if the case), needs to be treated accordingly to well-defined principles.

### 3.2. Types of Identity Management Solutions on Blockchain

In [6, 15], the authors classify DLT-based identity management solutions into two categories: *Self-Sovereign Identity* and *Decentralized Trusted Identity*.

**Self-Sovereign Identity (SSI)** is owned, controlled, and managed by the user, with no need to rely on a third party [6, 9]. In general, any SSI solution works as follows. The users generate their digital identities (one or several identifiers and attributes per user, depending on the case). The users add these identities to the blockchain, tied with a cryptographical public key, following the normal blockchain procedure [6]. The public key permits anyone to challenge and verify the authenticity of the user, as being the only one to possess the corresponding private key and hence can correctly respond to the challenge. More generally, any authentication procedure can be tied to the identity on the blockchain. However, usage of public-private key pairs is the most common method, and such SSI solutions are sometimes in the literature called *Decentralized Public Key Infrastructure* [9]. The users are then issuing *claims* of identities that need to be attested by other users to endorse their identity. These users can be family or friends that know and guarantee for one user's identity or institutions that guarantee on identities related to identity cards, passports, driving license, or other legal papers used to prove identity. They are called *claim-verifiers*, as they verify and validate the users' claim. The verifiable claims are managed outside of the blockchain, to preserve privacy, as sometimes issuer need to make public to the verifier private information to prove their link to the digital identity. Thus, a trustful relationship between the issuer and the verifier must be established before a verification, off-chain [9]. Based on what exactly is stored in the blockchain, there are two main SSI models: the *Identifier Registry Model* and the *Claim Registry Model*. The difference between them is that the second model does not only register identifiers in the blockchain, but also stores the cryptographic hashes for the identity associated claims [9]. The Identifier Registry Model is more privacy-preserving (claims are transparent for somebody reading the blockchain), however, it does not provide the same level of protection against tampering, as no claim registry could be used for additional verifications [9].

A system that implements SSI should comply with the *Ten Principles of Self-Sovereign Identity* [17]. They are an adaptation of the general principles, which concentrate more on the control of the end-user towards his own identity (control, consent, and access to his digital identity). These ten principles can be grouped into three main categories: *(1) security*, *(2) controllability*, and *(3) portability* [8, 9]. Security refers to *protection* (of users' privacy over the needs of the network), *minimalization* (of data exposed to respond to a specific action or task), and *persistence* (of the identifiers for a long time, depending on the users' needs). Controllability includes *existence* (of the entity that a digital identity corresponds to), *control* (of the digital identity by the user, who can update, hide or use his identity), and *consent* (of the users to use their identity). Portability deals with *interoperability* (across different platforms and services, at global level), *transparency* (in functionality, administration, and operation of the system) and *access* (of users to their complete data, with restrictions to access another users' information).

These principles are in accordance with the *laws of identity* and GDPR's principles previously presented in Subsection 2.1. Main security and privacy features such as protecting sensitive data by minimalization of exposure and limitation to the purpose needs with a strong focus on confidentiality is common to all sets of principles.

**Decentralized Trusted Identity (DTI)** assumes a service that proves the user's identity and records his digital identity on the blockchain. Proving of the user's identity relies on a general trusted method or (inter)national IDs such as the passport and requires the presence of an external authority [6]. Same as for SSI, additional identification attributes can be further tied to the digital identity. Note that, in fact, the existence of a trusted authority centralizes (to some extent) the users' enrollment in the blockchain: the service that performs the identity verification is usually proprietary and might sometimes be seen as a single point of trust (at least locally, where only specific entities might be the only one in charge with the verification and publishing of digital identities in the blockchain).

Both SSI and DTI identity management solutions can be built on top of either permissionless or permissioned blockchains. The type of the blockchain used to implement the solution has direct implications on the properties of the identity management solution. We will discuss this for the implementations below.

## 3.3 Existing implementations

We describe several blockchain-based identity management implementations, with focus on privacy and security aspects. Table 2 gives an overview of the examined solutions, divided by type (SSI, respectively DTI). For each implementation, we also indicate the type of blockchain (permissioned or permissionless) that can be used with it.

Table 2. Classification of implementations based on blockchain

| Implementation | Type (SSI/DTI) | Open-Source | Blockchain Type (Permissionless / Permissioned) | Blockchain Implementation |
|---|---|---|---|---|
| **Namecoin [23]** | SSI | yes | permissionless | Bitcoin fork |
| **Blockstack [24]** | SSI | yes | permissionless | Blockstack (Bitcoin) |
| **uPort [25]** | SSI | yes | permissionless | Ethereum |
| **Sovrin [26]** | SSI | yes | permissioned | Sovrin Network |
| **EverID [27]** | SSI | no | permissioned | Ethereum private |
| **SelfKey [28]** | SSI | yes (beta) | permissionless | Ethereum |
| **ShoCard [29]** | DTI | no | permissionless / permissioned | (Bitcoin) |
| **Sora [30]** | SSI | (under development) | permissioned | Hyperledger Iroha |
| **LifeID [31]** | SSI | yes | permissionless | (Ethereum) |
| **IDchainZ [32]** | DTI | (prototype) | - | ChainZy Smart Ledger |

*Namecoin* [23] was the first step towards using blockchain for identity management systems. Namecoin is considered the longest surviving fork of Bitcoin, and it binds human-readable names with IP addresses, in the sense of a Domain Naming System (DNS) [10, 23]. Hence, Namecoin is, in fact, a *naming system* and not an identity management system but we refer to it for historical reasons. Currently, Namecoin is an experimental open-source technology that aims to provide several features: attach attributes (e.g., Pretty Good Privacy (PGP) keys, e-mail) to a digital identity, decentralize Transport Layer Security (TLS) certificate validation, access websites using the *.bit* top-level domain that uses Bitcoin to decentralize website addresses [23]. Concerning identities, Namecoin allows fetching data associated with an identity in JavaScript Object Notation (JSON) format. The exact format, as well as the set of attributes that can be associated with an identity are available at [43]. This raises some privacy concerns, as anyone can access the data. Note that Namecoin is derived from Bitcoin, so it is permissionless, allowing anyone to join the process and read the information stored in the blockchain. Namecoin was proved vulnerable to the 51% attacks against blockchain [44], and more recently, it has been shown that an adversary can take ownership of any *.bit* domain [45].

*Blockstack* [24, 44] is an open source solution that extends Namecoin. Motivated by the possibility of a 51% attack against Namecoin, Blockstack brings a novel contribution: it allows migration to another blockchain in case of a major attack. This is possible by defining several layers, with the actual blockchain being the first layer that runs under a *virtual chain* logical layer that allows high flexibility [44]. Unlike Namecoin, it uses encryption to protect users' data, giving more control to the user using public cryptography (i.e., the user uses his private key for decrypting and signing the data).

*uPort* [25] is an open-source identity management system that claims to provide the users with a self-sovereign identity registered on the Ethereum blockchain using a mobile application. The uPort identifier is, in fact, an address on the Ethereum blockchain, while the user public key resolution and the management of the identifiers are done by using a smart contract (currently called *EthereumDIDRegistry*). The user can control the changes of his data by signing with the corresponding private address. Any entity can query the *EthereumDIDRegistry* and therefore, even if the data itself might be encrypted, the attributes structure and other metadata can leak sensitive information [6]. Moreover, because the user has no full control on what other entities can query, it means it has no full control over his attributes' disclosure, which makes it susceptible to not being a complete SSI solution [7]. uPort provides recovery mechanisms for the private key, which is stored on the user's mobile: either recovering the identity by using the seed words (from which the private key is derived) or by using a group of trustees previously chosen. The first method implies advanced user management skills, while the second one might be open to attacks such as coalitions of malicious trustees (the IDs of the trustees might be linked to the victim), or to direct replacement of trustees [6, 7]. The users can temporarily delegate his identity to other users and they can perform actions on behalf of the owner. There is also a chance that the uPort identity management system has unusable accounts due to loss of private key without naming trustees [7]. uPort also lacks portability in the sense that only internal identities can attest claims of other uPort identities [16].

*Sovrin* [26, 46] is an open-source, public service specially designed for identity management on blockchain. It runs on a permissioned blockchain, where only validated nodes take part in the consensus protocol, while anyone can use it for transactions. The use of a permissioned blockchain makes Sovrin take advantage of increased efficiency in reaching consensus, improved transaction rate and be less susceptible to 51% attacks [6]. The Sovrin blockchain is closely related to the Hyperledger Indy, a blockchain that is now under the Linux Foundation and Evernym, a platform dedicated to services and products based on Sovrin identity [47]. To preserve minimization of data exposure, Sovrin makes use of ZKP for all verifiable identity claims and they claim that it is possible to share selective attributes connected to the identity credentials without disclosing the credentials. Moreover, Sovrin tries to reduce correlation by separating the data from the identifiers, making a linkage to an identity difficult without additional information stored separately [46]. Similar to uPort, it provides a mobile application and accepts key recovery by means of trustees [6, 26].

*EverID* [27, 48] is an identity management system that runs on a private Ethereum instance hosted on EverID operated hardware. It is part of a larger solution, called Everest [27]. It uses proprietary datagrams to store users' identity information that are further cryptographically secured in some storage arrays. It is not clear to what extent the proprietary algorithms include cryptographical aspects as well. If non-open, proprietary cryptographical mechanisms are used, then the product breaks the well-known *Kerckhoffs principle*. The user's data is protected by several mechanisms such as public-private key pair, biometric means (face and fingerprint recognition), password and Personal Identification Number (PIN) [48]. Usage of so many authentication means should be properly tested, both from a security and a usability perspective. As a feature, in EverID the digital identity can be stored in the cloud and therefore its persistence is not linked to a physical device. On the weaknesses side, for claim verifications, the user has to disclose full data, which overcomes the principle of minimal exposure of data [16].

*SelfKey* [28, 49] is an open-source identity management system that runs on Ethereum public blockchain. For individual users, the data is stored on the user's device, being in control of the owner. Only if approved, other entities can access specific data. In the 2017 whitepaper, SelfKey planned to use uPort for recoverability of lost private keys. If so, then all vulnerabilities of uPort remain valid in the context of SelfKey too. Although in the literature SelfKey is supposed to have implemented ZKP to minimize the exposure of data [16], we found no official evidence of that. The SelfKey whitepaper mentions ZKP as a possibility, keeping in mind its overload.

*ShoCard* [29, 50] is an identity management platform that can use any permissioned or permissionless blockchain. As an advantage over solutions such as Blockstack, ShoCard supports not only migration but also multiple types of blockchain at the same time [7]. This might come as a response to a timing problem, as ShoCard operating on Blockchain might expect long processing times [6]. On the other hand, ShoCard is not open source and it is centralized in the sense that all records written on the blockchain pass via a ShoCard server. The ShoCard server does not break confidentiality directly, as the messages are encrypted under a public key (in a *secure envelope*) but restricts decentralization and maintains the availability of the solution for the lifetime of the ShoCard company only [6]. For self-certification, the user needs a mobile device that collects the name and value of his identity. The identity might be a phone number, an e-mail address, a scan of a valid document (e.g., passport, driving license), or biometric data (e.g., iris-scan, facial image, voice). To protect leakage, the names and values are locally stored apart, and the solution only publishes in the blockchain the salted hash, signed under the user's public key [50].

***Sora*** [30, 51] is an identity solution based on the Iroha permissioned blockchain. Both Sora and Iroha are currently under development. The Sora mobile app allows a user to generate a pair of cryptographic keys and store a salted hash of his private data in the blockchain. The solution basically follows the model of a general SSI solution. To prevent key lost, the public-private keys pair is stored on a central server in an encrypted form [30]. Encryption is performed under a key derived from a 8-digits password, which must satisfy some security requirements. We note two shortcomings here. First, if the storage of the keys is centralized, the solution cannot aim for full decentralization. Second, the security of the solution resides in the security of the master password, which is prone to human selection and hence might be vulnerable to known attacks, such as dictionary attacks.

***LifeID*** [31, 52] is an open-source identity management solution designed to work on any permissionless blockchain capable of smart contracts (e.g., Ethereum). It allows the implementation of ZKP to minimize sensitive data exposure. LifeID presents three key recovery procedures: self-backup, trusted organization backup, and backup using a trusted group of individuals (e.g., friends, families). LifeID makes no use of passwords but uses biometric authentication instead. Hence, it requires a mobile with biometric capabilities. The current status of the project is not clear, as the official website posted no updates since last year.

***IDchainZ*** [32] is a proof of concept DTI, built on top of the ChainZy Smart Ledger. The system has two distinct mutual distributed ledgers – one for holding the individually encrypted documents and the other a transaction ledger that holds the keys of the documents [32]. There is not enough public information available on the website in order to conduct an in-depth security evaluation of the proposal. We found no specification about the type of blockchain (permissioned or permissionless) either.

Other existing implementations include ***BitID*** [53], which is not a general solution, but it is a specific Bitcoin authentication protocol that requires a Bitcoin address. Moreover, other digital identity solutions on blockchain or standards are being implemented or considered by organizations and startups, e.g., ***Bitnation*** [54], ***ID2020*** [55], ***Ethereum Identity Standard ERC-725/735*** [56], ***W3C Decentralized Identifiers*** (DIDs) [57], ***Civic*** [58], ***Jolocom*** [59], ***WiSeID*** [60].

## 4. DISCUSSION

Blockchain-based identity management solution was proposed as a candidate solution to the identity management problem. The main property a blockchain should have by definition – *decentralization* – could increase freedom of choice and independence of big companies and organizations. It could become a strong alternative to the existing identification and authentication systems widely used (e.g., via Google or Facebook) [44]. Moreover, SSI solutions give, in theory, full control to the user with respect to his data, decreasing the danger of identity theft. However, important questions persist: *Does identity management on blockchain bring real benefits?*, *Are the benefits brought by identity management on blockchain strong enough to overcome the complexity they introduce?* and, most of all, *Is blockchain a real solution to identity management problems, or identity management is just another artificial context to make use of "the most over-hyped technology in human history"* [61]?

Table 3. Claimed properties of the analyzed blockchain implementations

| Implementation | ID Secure Verification | Long-term Validity | ID Management (recovery of identity / private key) | ID Self-management (e.g., set attributes, delegation) |
|---|---|---|---|---|
| **Namecoin [16]** | no | no | no | set attributes |
| **Blockstack [18]** | yes | yes | 12-word recovery phrase | no |
| **uPort [21]** | yes | yes | 12-word recovery phrase, group of trustees | set attributes, delegation |
| **Sovrin [22]** | yes | yes | "social recovery" (trustees having shards of data) | no |
| **EverID [23]** | yes | yes | mnemonic phrase, biometrics, PIN, password | set attributes |
| **SelfKey [24]** | yes | yes | no, but planned | no, partially planned (delegation) |
| **ShoCard [25]** | yes | yes | three-factor recovery process (from phone number, e-mail, phrase, ID, scanned document) | set attributes |
| **Sora [26]** | yes | yes | not provided | set attributes |
| **lifeID [27]** | yes | yes | 12/24-word recovery phrase, group of trustees, trusted organization | set attributes |
| **IDchainZ [28]** | not provided | yes | not provided | not provided |

Despite the fact that there is no straight positive answer to the questions above, blockchain-based identity management solutions have been implemented, and people started to use them at personal, organizational and governmental levels. An example is the pilot implementation of *uPort* to manage the digital identity of the citizens of the Swizz city Zug, with possibility to use it for e-voting in the future [62].

Currently, blockchain-based identity management solutions are still emerging, and more research must be performed to gain a better knowledge of their functionalities. Table 3 summarizes some important aspects of the implementations described in Section 3.2, for what the solutions claim to offer. We further indicate more directions that must be considered for in-depth analysis and future research.

***Specific aspects of blockchain-based identity management*** that should be considered can be either particular to identity management or inherited from the underlying blockchain technology.

In the first category, we mention the verification of the real identity. It remains open how this can be performed securely in totally decentralized environments (because of e.g., 51% attack, malicious trustees, or fake identities created by combinations of different attributes of real entities [63]). Moreover, decentralization itself can be an issue. Not all implementations are fully decentralized (e.g., ShoCard – see Subsection 3.3), and even if they are, then there is a great need for trust [6]: trust between the claimers and the verifiers, trust in the majority of participants, etc. Even more, the decentralization of the underlying blockchains is also debatable [64]. Furthermore, attention must be paid to possible faults. Key recovery mechanisms should be available in case of lost private keys, to prevent identity loss. evertheless, they must be implemented with care. Solutions such as recovery via trustees may be exposed to vulnerabilities (e.g., uPort, Sovrin - see Subsection 3.3), and private key recovery from passwords (e.g., Sora – see Subsection 3.3) is strongly not recommended. Many solutions are built for mobile environments, and some store the private key directly on the smartphone [9]. This transforms the smartphone in a single point of vulnerability from a user perspective. Finding appropriate key recovery solutions is a research topic.

By definition, blockchain should be immutable (i.e., published information is unchangeable), long-preserving (i.e., information remains in the blockchain for at least the life-time of the blockchain) and, for permissionless blockchains, data is public. While this transparency can be good in some contexts, it can harm privacy. Attributes should not be published in the blockchain, at least not in cleartext. Moreover, even if data is encrypted, metadata might leak sensitive information (e.g., uPort – see Subsection 3.3). Nevertheless, pattern analysis of on-chain data and exchanged messages is a general risk to information disclosure [35].

In the second category, we note that in many situations the underlying blockchain immutability does not hold, e.g., in case of 51% attacks on Proof of Work (PoW). A financial incentive can be enough to mount such an attack, if the rewards are high [15]. According to [65], investment in computational mining power to equal the total mining power in Bitcoin is 400 million dollars, while the financial gain over the network can be estimated to much more. High electricity consumption [15] and performance should also be considered when thinking about blockchains as an underlying technology. Possible solutions to problems on the underlying blockchain are migration (e.g., Blockstack – see Subsection 3.3) or multiple-blockchain usage (e.g., ShoCard – see Subsection 3.3). However, some problems appear from the definition of blockchain and cannot be easily overcome. An example is the infinite ledger problem: the length of the blockchain is endlessly increasing, which introduces difficulties for storage, at least for full new nodes (e.g., the download time for Bitcoin is currently up to three days) [44]. It remains open how would identity management solve this concerning scalability in time and users.

*Cryptographical aspects* have a direct impact on the privacy and security of identity management on blockchain. Use of proprietary, not public solutions that break Kerckhoffs principle (e.g., EverID – see Subsection 3.3) and passwords (e.g., Sora – see Subsection 3.3) that might be vulnerable to well-known dictionary attacks are examples of bad practice. Moreover, if attributes or other sensitive data are published in the blockchain, future advances of cryptanalysis might damage privacy. Hash functions and public key encryption used to protect data are just computationally infeasible, so in time they might get broken. Additionally, most of the existing public key solutions are known to be vulnerable to quantum attacks, so in the future, quantum-resistant primitives must be accommodated in the blockchain-based identity management solutions. How can this be done and to what extent it will influence the efficiency and functionality is a research direction. Similar, the usage of ZKP for proving identity claims might be a good option from a cryptographic point of view, but the tradeoffs in complexity need to be analyzed in more depth. Lastly, effective key management remains a challenge in cryptographic terms too [6].

*Usability aspects* are important for a successful identity management system. However, research in usability and user experience seems to be in an incipient stage. Are end-users and developers willing to use such solutions for the long term? For example, are end-users able to securely manage their identifiers and credentials by themselves? They could (partially) delegate control for certain periods of time or rely on services such as recovery mechanisms in case of loss [35]. uPort mentions a list of requirements to onboard new users and admits that it is not a trivial task [66], and EverID introduces many means of authentication (password, PIN, biometric recognition), making the solution non-user-friendly (see EverID – Subsection 3.3). Demographic aspects make users behave differently and more research is needed in this area [67]. End-users should be aware of the implications of using blockchain-based solutions and to what extent they can satisfy their needs. But many times, complex implementation details and lack of explanations from solution providers make users unaware of the risks they are exposed to. One inconvenient of PKI is its complexity, but blockchain-based solutions are also complex, so they might not be a good replacement from this perspective.

## 5. CONCLUSIONS

This paper discusses identity management on blockchain. The adoption of the blockchain technology to solve any type of problem must be overcome by proper analysis of benefits and tradeoffs. We have shown that existing implementations have flaws and more analysis needs to be conducted to overcome different challenges. Further investigation needs to be done, and the real benefits of blockchain-based identity management in relation to the complexity of usage, implementation, and maintenance must be carefully taken into consideration.

## ACKNOLEDGEMENT

This work was partially supported by a grant of Romanian Ministry of Research and Innovation CCCDI-UEFISCDI project no. 17PCCDI/2018.

---

[2] Note: all the links were last accessed in April 2020